\documentstyle[aps,prl,floats,twocolumn,epsf]{revtex}

\begin{document}

\noindent {\bf Comment on ``Density Matrix Renormalization Group
Study of the Haldane Phase in Random One-Dimensional
Antiferromagnets" } In a recent Letter\cite{hida}, Hida presented
numerical results indicating that the Haldane phase of the
Heisenberg antiferromagnetic (HAF) spin-1 chain is stable against
bond randomness, for box distributions of the bond strength, even
when the box distribution stretches to zero bond strength. The
author thus concluded that the Haldane phase is stable against
bond randomness for any distribution of the bond strength, no
matter how broad.
This conclusion contradicts our earlier prediction\cite{hyman}
that a sufficiently broad bond distribution drives the HAF spin-1
chain through a critical point into the Random Singlet phase. In
this Comment, we (i) point out that the randomness distributions
studied in Ref. \cite{hida} do {\em not} represent the broadest
possible distributions, and therefore these numerical results do
{\em not} lead to the conclusion that the Haldane phase is stable
against {\em any} randomness; and (ii) provide a semiquantitative
estimate of the critical randomness beyond which the Haldane phase
yields to the Random Singlet phase, in a specific class of random
distribution functions for the bond strength.

(i) The box distribution function for bond strength $J$ used in
Ref. \cite{hida} takes the form
$P_{box}(J)=1/W$
for $1-W/2\le J \le 1+W/2$, and $P_{box}(J)=0$ otherwise.
There is no randomness for $W=0$, and the randomness increases with
$W$. The maximum randomness for this class of distribution
functions is reached at $W=2$, as further increasing $W$ would
introduce {\em ferromagnetic} bonds to the system and change the
physics completely. On the other hand, the $W=2$ distribution does
{\em not} represent the broadest possible distribution among all
possible bond distributions. In fact, it is a member of the class
of power-law distributions:
$P_{pow}(J)=\alpha J^{-1+\alpha}$
for $0 \le J \le 1$, and $P_{pow}(J)=0$ otherwise. The case $W=2$
of $P_{box}$ corresponds to $\alpha=1$ in $P_{pow}$. The power-law
distributions are of particular interest and importance in studies
of random spin chains, because under real space renormalization
group (RSRG) transformations, all generic distribution functions
will flow to power-law distributions in a HAF spin-1/2
chain\cite{dm,fisher1}.
The broadness of a power-law distribution is parameterized by the
single positive exponent $\alpha
> 0$. The smaller $\alpha$ is, the broader the distribution is on
a logarithmic scale (the width is $1/\alpha$)\cite{fisher1}.
Clearly, for any $0 < \alpha < 1$, the distribution function $P_{pow}(J)$
is broader than those studied in Ref. \cite{hida}. Thus the
numerical results of Ref. \cite{hida} do {\em not} necessarily
lead to the conclusion that the Haldane phase in HAF spin-1 chain
is stable against any bond randomness, no matter how broad the
distribution\cite{note}.

Hida also argues that the persistence of the string order is a
consequence of the absence of translation symmetry in the
effective spin-1/2 description of the random HAF spin-1
chain\cite{hyman}, similar to what happens in the random dimerized
spin-1/2 chain\cite{hyman2}. We point out that the Random Singlet phase in the
effective spin-1/2 model\cite{hyman}
does not have translational symmetry, and that
broken translational symmetry and string order are not synonymous.

(ii) While a naive extension of the original form of the
RSRG\cite{dm,fisher1} cannot be applied directly to the random HAF
spin-1 chain unless the initial distribution function is
sufficiently broad, we have developed a RG scheme that is suitable
for the random HAF spin-1 and other random spin chains. The key
ingredient of this scheme is to project out the highest energy
state(s) in the strongest bond and keep all lower energy states at
each RG step. Details of this scheme will be presented
elsewhere\cite{hy}. Using this scheme to simulate the HAF spin-1
chain with power law distributions, we have found two phases
separated by a critical point $\alpha_c \approx 0.67$. For $\alpha
> \alpha_c $ (which includes the flat distributions of Ref.
\cite{hida}) the system is in the Haldane phase characterized by
the string order, while for $\alpha < \alpha_c $ it is in the
Random Singlet phase. While there are approximations involved and
our estimate of $\alpha_c $ may not be very accurate numerically,
it clearly suggests that one needs bond distributions that are
{\em broader} than those studied\cite{hida} to destabilize the
Haldane phase.

K.Y. was supported by NSF DMR-9971541 and the Sloan Foundation.

\vspace{12pt} \noindent Kun Yang\\ NHMFL and Physics Department\\
Florida State University\\ Tallahassee, Florida 32310

\vspace{12pt} \noindent R. A. Hyman\\ Department of Physics\\
DePaul University\\ Chicago, IL 60614-3504


\begin{references}
\bibitem{note} For $0 < \alpha < 1$,
$P_{pow}(J)$ is singular at $J=0$. However this is not a problem
because the singularity is integrable, and the probability for
appearance of $J=0$ bonds is zero.
\end{references}
\end{document}